\documentclass[aps,showpacs,11pt,letterpaper,nofootinbib,amsmath,amssymb]{revtex4-2}%
\usepackage{latexsym}%
\usepackage{epsf}%
\usepackage{graphicx}%
\usepackage{epsfig}%
\usepackage{graphicx}%
\usepackage{epstopdf}%
\usepackage[dvips,dvipdfmx,hypertex]{hyperref}%

\def\cM{{\mathcal{M}}}

\def\hP{{\widehat{P}}}
\def\hM{{\widehat{M}}}

\def\cH{{\mathcal{H}}}

\def\cE{{\mathcal E}}

\DeclareMathAlphabet{\mathpzc}{OT1}{pzc}{m}{it}

\newcommand{\beq}{\begin{equation}}
\newcommand{\beqn}{\begin{equation}\nonumber}
\newcommand{\eeq}{\end{equation}}
\newcommand{\bea}{\begin{eqnarray}}
\newcommand{\bean}{\begin{eqnarray}\nonumber}
\newcommand{\eea}{\end{eqnarray}}
\newcommand{\ba}{\begin{align}}
\newcommand{\ea}{\end{align}}

\begin{document}

\title{Quantum Collapse of a Thin Shell Revisited}
\author{Cenalo Vaz\footnote{\tt Cenalo.Vaz@uc.edu}}
\affiliation{University of Cincinnati, Cincinnati, OH 45221-0011.}
\begin{abstract}
 
There are several possible choices of the time parameter for the canonical description of a self-gravitating thin shell, but 
quantum thories built on different time parameters lead to unitarily inequivalent descriptions. We compare the quantum collapse
of a thin dust shell in two different times {\it viz.,} the time coordinate in the interior of the shell (originally addressed
in \cite{hajicek92a}) and the time coordinate of the comoving observer (proper time). In each case, we obtain exact solutions
to the Wheeler-DeWitt equation requiring only a finite and well behaved $U(1)$ current. The two quantum theories are complementary
and each highlights the role played by the Planck mass: stationary states of positive energy in interior time exist only if
the shell rest mass in smaller than the Planck mass. In proper time they exist only when the shell rest mass is {\it greater}
than the Planck mass. In coordinate time there are both scattering states and bound states with a well defined energy spectrum.
In the proper time description there are only bound states, whose spectrum we determine.

\end{abstract}
\maketitle

\section{Introduction}

Many paradoxes associated with the formation of spacetime singularities seem to point to the need for a quantum theory of the
gravitational field, but this has proved to be a very difficult problem. Experimental study is hampered by the weakness of the
gravitational interaction with matter, therefore there has been virtually no experimental input for an informed theoretical
exploration of the problem and many conceptual issues remain unresolved. One of these is the unitary inequivalence of quantum 
theories built on different time coordinates. To get a handle on this, it is useful to examine a system in which exact solutions 
can be found in quantizations based on different time parameters. 

Often, progress is made by examining simplified models that capture some of the troublesome features of the full problem but largely 
avoid most of the technical difficulties. A spherical shell, in the limit in which the shell thickness is taken to be infinitesimal, 
is an example of a toy model that captures some of the essential features of gravitational collapse \cite{berezin88,martinez89,
farhi90,visser91,hajicek92b,berezin97,adler05,alberghi06,wang09,poisson19}. On the classical level, the shell has just one degree of 
freedom and is completely described by its radius, $R(t)$ and its conjugate momentum, $P(t)$. Yet, various versions of it form a rich 
enough collection of physical systems to describe the final stages of gravitational collapse, Hawking radiation and the formation 
(or avoidance) of gravitational singularities \cite{visser04,xu06,vachaspati08,vachaspati09,paranjape09,bardeen14,ziprick16,
binetruy18,baccetti20}.

A quantum dust shell (of vanishing surface tension) that is collapsing in a vacuum can be exactly solvable while incorporating the 
fully relativistic gravitational interaction with matter \cite{hajicek92a}. This apparent simplicity comes, however, with the problem 
of time \cite{kuchar92,isham93} mentioned earlier. It manifests itself as follows: because the shell dynamics are constructed by an 
application of the Israel-Darmois-Lanczos (IDL)\cite{israel66,darmois27,lanczos24} junction conditions, there are three distinct time 
variables present in the problem, each of which is ``natural'' in some setting. These are (i) the time coordinate appropriate to the 
interior of the shell, (ii) the time coordinate in the exterior of the shell and (iii) the comoving (proper) time of the shell. There 
is one conservation law that may be construed as a first integral of an equation of motion. At issue is the construction of a Hamiltonian 
for the system: the conservation law is obtained in terms of the dependent variables (velocities) of a canonical theory and, depending 
on which time variable is chosen, different Hamiltonians are obtained.

In this paper we compare exact quantizations of a dust shell in two different times, {\it viz.,} the coordinate time interior to the
shell and the shell's proper time. In section II, we briefly summarize (for completeness) the IDL formalism for the dust shell and
obtain the first integral of the shell's motion. If the exterior geometry is taken to be a vacuum spacetime, the first integral of
the motion involves two constants which are interpreted as the the rest mass, $m$, of the shell and the total (ADM) mass, $M$, defining
the exterior. Of these, $m$ is a constant over the entire phase space, whereas $M$ is a dynamical variable which represents the total
energy, $E$, of the system. Following \cite{hajicek92a} we take the ADM mass to generate the evolution in the time coordinate of the 
interior of the shell. We take this to be a canonical choice defining the system, not a just convenient trick. This then allows for 
the construction of an effective Largrangian for the system. Once it is known for one particular time variable, the effective Lagrangian 
may be re-expressed in terms of either of the other two time variables (Schwarzschild time in the exterior and proper time) and, from 
the effective Lagrangians, Hamiltonians for the evolution in all three times may be obtained. The proper time Hamiltonian obtained in this 
way is structurally identical to the Hamiltonian obtained in \cite{vaz01} for a dust ball in the LeMa\^ itre-Tolman-Bondi (LTB) 
\cite{ltb} collapse models by an application of a canonical chart analogous that that employed by Kucha\v r \cite{kuchar94,brown94} 
to describe the Schwarzschild black hole. The Hamiltonians obtained in this approach differ from those that would have been obtained 
had one not made the canonical choice of \cite{hajicek92a} at the start. 

The interior and exterior do not cover the entire spacetime, which is the union of the two with the 
shell as a boundary. But in the quantum theory, the shell is an ill defined boundary because the wave function is smeared over
all values of the radius and the terms ``interior'' and ``exterior'' lose their meaning. Therefore, a better quantum picture is 
likely obtained if the quantum evolution is examined in the shell's proper time. In section III, we analyze the quantum theory 
from the comoving observer's point of view. The Wheeler-DeWitt equation is an {\it elliptic} Klein-Gordon equation with a well
defined positive semi-definite inner product for energies less than the shell mass. Here we show that no stationary states with a well
behaved $U(1)$ current exist if the mass of the shell is less than the Planck mass. When the shell rest mass is greater than the
Planck mass only bound states exist and we find their energy spectrum.

We compare the results of the quantization in proper time and interior time in section IV. The quantum description by the 
comoving observer is, in some sense, complementary to the description by the interior observer. From the interior observer's point 
of view, no solutions exist when the shell mass is {\it greater} than the Planck mass and there are both scattering states and bound 
states otherwise. If this or an analogous limitation on the mass were to hold true for {\it thick} shells then one could clearly 
eliminate the quantization based on interior time as contradicting observation, but this is yet an open question. To build the Hilbert 
space and obtain the energy spectrum for the thin shell it is only necessary to require that a lowest energy state exists and that 
the $U(1)$ current is well behaved and finite everywhere.  We conclude in section V with a brief summary and outlook.

\section{Classical Thin Shells}

The equation of motion of a spherical, thin, massive shell is obtained by applying the Israel-Darmois-Lanczos conditions on the
timelike surface $\Sigma = \mathbb{R}\times \mathbb{S}^2$ that represents its world sheet. The world sheet forms the three
dimensional boundary between an internal spacetime, $\cM^-$, and an external spacetime, $\cM^+$. $\cM^{\mp}$ are described in
coordinates $x_\mp^\mu$ by metrics $g_{\mu\nu}^\mp$ that solve Einstein's equations. Let $\xi^a$ be a set of intrinsic coordinates
on the surface of the shell and differentiable functions of $x^\mu_\mp$, then ${e^\mp}^\mu_a = \partial x_\mp^\mu/\partial \xi^a$
are the components of the three basis vectors on this surface and $h^\mp_{ab} = g^{\mp}_{\mu\nu} {e^\mp}^\mu_a{e^\mp}^\nu_b$ is the induced metric on the shell on the two sides of it. The first junction condition
requires the shell to have a well defined metric, {\it i.e.,} $h^-_{ab} = h^+_{ab}$.

Let $n^\mp_\mu$ represent the unit outward normal to the shell ($n^2 = +1$ for a timelike surface and $n^\mp_\mu {e^\mp}^\mu_a = 0$) 
and $K^\mp_{ab}$ the extrinsic curvature on either side of it,
\beq
K^\mp_{ab} = {e^\mp}^\mu_a {e^\mp}^\nu_b \nabla^\mp_\mu n^\mp_\nu.
\eeq
If $\kappa_{ab} = [K_{ab}]= K^+_{ab}-K^-_{ab}$, the second junction condition, which follows from Einstein's equations,
says that the surface stress energy tensor, $S_{ab}$, of the shell is given by 
\beq
S_{ab} = -\frac\varepsilon{8\pi}\left(\kappa_{ab} - \kappa h_{ab}\right)
\label{shellstress}
\eeq
where $\kappa = \kappa^a_a$ and $\varepsilon = +1$ for a timelike shell.

If $\cM^\mp$ are taken to be vacuum spacetimes, then spherical symmetry implies that $g_{\mu\nu}^\mp$ are Schwarzschild metrics,
with mass parameters $M^\mp$ respectively, and $M^+$ represents the total mass of the system. We may write the respective line
elements as
\beq
ds^2_\mp = - g^\mp_{\mu\nu} d x_\mp^\mu d x_\mp^\nu = B^\mp dt_\mp^2 - \frac 1{B^\mp} dr^2_\mp - r_\mp^2 d\Omega^2
\eeq
where $B^\mp = 1-2GM^\mp/r_\mp$ and we have assumed that the interior and exterior share the same spherical coordinates, $\theta$ 
and $\phi$. The shell is described by the parametric equations $r_\mp = r = R(\tau)$, $t_\mp = t_\mp(\tau)$, where $\tau$ is the 
proper time for comoving observers and the interior and exterior time coordinates are related to the shell proper time (and
indirectly to each other) by
\beq
\frac{dt_\mp}{d\tau} = \frac{\sqrt{B^\mp + R_\tau^2}}{B^\mp},
\label{timeders}
\eeq
where the subscript indicates a derivative with respect to $\tau$. Choosing the intrinsic coordinates of the shell to be $\xi^a =
\{\tau,\theta,\phi\}$, the induced metric is
\beq
ds_\Sigma^2 = d\tau^2 - R^2(\tau) d\Omega^2.
\eeq
Again, from the normals on either side of the shell,
\beq
n_\mu^\mp = \langle -\dot R, \frac{\sqrt{B^\mp + R_\tau^2}}{B^\mp},0,0\rangle,
\eeq
the non-vanishing components of the extrinsic curvature are given as
\beq
{K_\mp^\theta}_\theta = {K_\mp^\phi}_\phi = \frac{\beta^\mp}R,~~ {K_\mp^\tau}_\tau = \frac{\beta_\tau^\mp}{\dot R_\tau},
\eeq
where
\beq
\beta^\mp = \sqrt{B^\mp + \dot R_\tau^2}.
\eeq
Therefore, according to \eqref{shellstress}, 
\bea
{S^\tau}_\tau &=& \frac{\beta^+-\beta^-}{4\pi GR} = -\sigma\cr\cr
{S^\theta}_\theta = {S^\phi}_\phi &=& \frac{\beta^+-\beta^-}{8\pi GR} + \frac{\beta_\tau^+ - \beta_\tau^-}{8\pi G R_\tau} 
= p
\label{shellstress2}
\eea
where we have set ${S^a}_b =\text{diag}(-\sigma,p,p)$. $\sigma$ represents the mass density of the shell and $p$ the pressure,
which, for dust shells, we take to be zero. 

Integrating the second equation in \eqref{shellstress2},
\beq
\beta^+-\beta^- = -\frac {Gm}R
\label{deltabeta}
\eeq
where $m$ is a constant of the integration, which represents the rest mass of the shell, as is seen by inserting this solution 
into the first. Equation \eqref{deltabeta} may be put in the form
\beq
M^+-M^- = \Delta M = m\sqrt{B^-+R_\tau^2} -\frac{Gm^2}{2R}.
\label{genconst}
\eeq
In this expression, $M^+$ is a dynamical variable whereas $m$ and $M^-$ are prescribed constants.

For a shell collapsing in a vacuum, $M^-=0$, $M^+ = M$. We relabel the time coordinate in the interior as $T~ (=t_-)$ and in the 
exterior as $t~ (=t_+)$, then 
\beq
M = m\sqrt{1+R_\tau^2} - \frac{Gm^2}{2R}
\label{vacuumconst}
\eeq
and using \eqref{timeders}
\beq
\frac{dT}{d\tau} = \sqrt{1 + R_\tau^2},~~ \frac{dt}{d\tau} = \frac{\sqrt{B + R_\tau^2}}B.
\label{timeders2}
\eeq
It is reasonable think of \eqref{vacuumconst} as a first integral of the motion and associate the ADM mass with the total energy, $E$,\
of the shell. There is a turning point in the shell motion (the shell is bound) so long as $m>E$, 
otherwise its motion is unbounded.

The right had side of \eqref{vacuumconst}, when expressed in terms of the momentum conjugate to $R(\tau)$, will then
represent the Hamiltonian for the evolution of the system, but it is given in terms of what are ``dependent'' variables in the 
canonical theory. The question is: in which of the three available time coordinates
is this Hamiltonian evolving the system? For example, if the evolution is taken to be in the shell proper time and the energy 
is taken to be $M$, the Hamiltonian is \cite{berezin88}
\beq
H = m \cosh \frac pm - \frac{Gm^2}{2R}
\label{hyperbolicham}
\eeq
The corresponding operator has derivatives of all orders, but it was shown to possess a positive self adjoint extension 
if the rest mass is less than the Planck mass \cite{hajicek92b}. On the other hand, if the the ADM mass evolves the system in the time 
variable of the interior of the shell ($T$) and \eqref{vacuumconst} is treated as
\beq
M = \frac m{\sqrt{1-R_T^2}} - \frac{Gm^2}{2R}
\label{vacuumconstT}
\eeq
where we have used \eqref{timeders2}, then one finds \cite{hajicek92a}  
\beq 
H = -p_{(T)} = \sqrt{p^2+m^2} -\frac{Gm^2}{2R}.
\label{haminttime}
\eeq
The equations of motion that follow from \eqref{haminttime} are derivable from the superhamiltonian
\beq
h_T = (p_{(T)} - f)^2 - p^2 -m^2 = 0,
\label{superhT}
\eeq
where $f(R) = Gm^2/2R$ is the shell self interaction, which is formally equivalent to the classical field equation of 
a charged scalar in a radial Coulomb potential. The shell quantum mechanics, built on this superhamiltonian, was studied in 
\cite{hajicek92a} subject to boundary conditions appropriate to its interpretation as a classical field theory.

One could also, in principle, imagine that it is preferable to describe the evolution in the external time \cite{vachaspati08,vachaspati09}, 
{\it i.e.,} from the standpoint of the asymptotic observer. Rewriting the constraint in terms of $R_t$, using \eqref{timeders2}, one gets
\beq
M = m\sqrt{1+\frac{B R_t^2}{B^{2}-R_t^2}}-\frac{Gm^2}{2R},
\label{vacuumconstt}
\eeq
but now the constraint involves the ADM mass on both sides and it is difficult to determine from it a Hamiltonian for the evolution of 
the system. The authors of \cite{vachaspati08} suggested that one should instead treat \eqref{haminttime} as the canonical Hamiltonian 
for the evolution in the internal time coordinate, $T$, and construct an effective Lagrangian, which is easily found to 
be
\beq
L = p R_T - H = -m \sqrt{1-R_T^2} + \frac{Gm^2}{2R}
\eeq
The effective action can now be transformed using \eqref{timeders}
\beq
S = -m \int dT \left[\sqrt{1-R_T^2} + \frac{Gm^2}{2R}\right] = -m \int dt\sqrt{B}\left[\sqrt{1-\frac{R_t^2}{B^2}}-\frac{Gm}{2R}
\sqrt{1-(1-B)\frac{R_t^2}{B^2}}\right]
\eeq
and from the transformed action, a new generalized momentum and Hamiltonian can, in principle, be obtained. At this point the 
ADM mass is no longer treated as a dynamical variable but as a global constant, on the same footing as the shell mass. Unfortunately,
it is difficult to extract the momentum from the above action, but one sees that the energy, expressed in terms of $R_t$, is 
\beq
E = m B^{3/2} \left[\frac 1{\sqrt{B^{2}-R_t^2}}- \frac{Gm}{2R\sqrt{B^2-(1-B)R_t^2}}\right]
\eeq
and differs considerably from the ADM mass in \eqref{vacuumconstt}. This system is technically difficult to analyze and exact solutions
cannot be obtained, so we will not pursue it further here. It was studied in the near horizon limit in \cite{vachaspati08}.

Similarly, transforming the action to proper time with the help of \eqref{timeders2},
\beq
S = \int d\tau \left[-m + \frac{Gm^2}{2R} \sqrt{1+R_\tau^2}\right]
\label{actiontau}
\eeq
one derives the Hamiltonian for the evolution in proper time,
\beq
\cH = - P_{(\tau)} = m - \sqrt{f^2-P^2}
\label{hamintau}
\eeq
where $P$ is the momentum conjugate to $R$,
\beq
P = \frac{f R_\tau}{\sqrt{1+R_\tau^2}},
\label{radialP}
\eeq
and we have set $f(R) = G m^2/2R$ as before. The proper energy is bounded from above
by the shell mass and the proper momentum is bounded above by $f$. As a result, the shell is always bound to the center. The
Hamiltonian is no longer a hyperbolic function of the momentum as in \eqref{hyperbolicham}, and the equations of motion that follow
from \eqref{actiontau} are generated by the superhamiltonian
\beq
h = (P_{(\tau)} + m)^2 + P^2 - f^2 = 0.
\label{superh1}
\eeq
It is surprisingly similar in structure to the superhamiltonian obtained in \cite{vaz01} for a marginally bound dust ball in a 
midisuperspace quantization of the Einstein-Dust system \cite{ltb}. As a midisuperspace problem, there are ambiguities 
associated with the construction of diffeomorphism invariant states in the quantization program. No such ambiguity appears in this 
minisuperspace problem, so the shell provides an excellent toy version of that problem. That said, there are some significant 
differences as well. The dust shells in a dust ball do not possess the self interaction represented by $f(R)$, and the interior of 
each shell is not a vaccum but a collection of dust shells, which provide the gravitational attraction to the center (we return 
to this in the concluding section).

\section{Shell Quantum Mechanics in Proper Time}

We will work with the superhamiltonian \eqref{superh1} and later compare the results with the quantization of \eqref{superhT} in the
next section. To get the wave equation, we follow Dirac and elevate the momenta to operators in the usual way. The structure of the 
superhamiltonian indicates that the DeWitt metric is $\gamma^{ij} = \text{diag}(1,1)$. Consequently, we choose the trivial measure, 
``$dR$'' and a factor ordering that is symmetric with respect $dR$. The Wheeler-DeWitt equation,
\beq
\left[(-i \partial_\tau + m)^2 - \partial_R^2 - f^2\right]\Psi(\tau,R) = 0.
\label{KGtau}
\eeq
is formally an {\it elliptic} Klein-Gordon equation of a particle moving in the potential $f^2 = G^2 m^4/4R^2$. Let us show that,
in the classical limit, \eqref{KGtau} yields the classical dynamical equations that follow from \eqref{hamintau}. Taking $\psi(\tau,R) =
e^{iS(\tau,R)}$ we find, to order $\hbar^0$, the Hamilton-Jacobi equation
\beq
\left(\frac{\partial S}{\partial\tau}+m\right)^2 + \left(\frac{\partial S}{\partial R}\right)^2 - f^2 = 0,
\eeq
whose solution may be given in the form
\beq
S(\tau,R) = - \cE \tau \pm \int dR \sqrt{f^2-(m-\cE)^2}.
\eeq
By the principle of constructive interference,
\beq
\frac{\partial S}{\partial \cE} = 0 = -\tau \pm \int \frac{dR(m-\cE)}{\sqrt{f^2-(m-\cE)^2}}.
\label{constrint}
\eeq
The functions
\beq
\frac{\partial S}{\partial R} = P = \sqrt{f^2-(m-\cE)^2}
\label{pHJ}
\eeq
and $R(\tau)$ defined by \eqref{constrint} should satisfy the equations of motion based on the Hamiltonian in \eqref{hamintau}.
Taking a derivative of \eqref{constrint} with respect to $\tau$,
\beq
1 = \pm \frac{(m-\cE)R_\tau}{\sqrt{f^2-(m-\cE)^2}}~~ \Rightarrow~~ m-\cE = \frac{f}{\sqrt{1+R_\tau^2}}
\label{mcE}
\eeq
which implies that
\beq
P = \frac{fR_\tau}{\sqrt{1+R_\tau^2}}~~ \Rightarrow~~ R_\tau = \frac P{\sqrt{f^2-P^2}} = \left\{R,\cH\right\}.
\eeq
Again, taking a derivative of $P$ in \eqref{pHJ} results in
\beq
P_\tau = \frac{ff'R_\tau}{\sqrt{f^2-(m-\cE)^2}} = \frac{ff'}{\sqrt{f^2-P^2}} = \left\{P,\cH\right\}
\eeq
where we have used \eqref{mcE} in the second step above. Therefore, the trajectories implied
by the principle of constructive interference in \eqref{constrint} are identical to those determined by the Hamiltonian
equations of motion that follow from \eqref{hamintau}.

For any two solutions of the wave equation, $\Phi$ and $\Psi$, there is a conserved bilinear current density,
\beq
J_i = - \frac i2 \Phi^* \overleftrightarrow{\partial_i}\Psi + m \delta_{i\tau}\Phi^*\Psi,~~ i \in \{\tau,R\},
\label{conscurtau}
\eeq
the time component of which specifies a physical inner product
\beq
\langle \Phi,\Psi\rangle = \int_0^\infty dR\left[- \frac i2 \Phi^* \overleftrightarrow{\partial_\tau}\Psi + m \Phi^*\Psi\right],
\label{inprod}
\eeq
sometimes referred to as the ``charge'' form in analogy with the classical charged field. Here, the charge form is positive semi-definite 
as long as $\cE<m$ and may be taken to represent a probability density. Therefore, with \eqref{inprod} we obtain an inner product space 
that can be extended to a separable Hilbert space by Cauchy completion \cite{reed80,wald94}. We confine our attention to stationary states,
\beq
\Psi(\tau,R) = e^{-i\cE \tau}\psi(R),
\label{statstate}
\eeq
which leads to the following radial equation:
\beq
\psi''(R) - \left[(m-\cE)^2 - \frac{\mu^4}{4R^2} \right]\psi(R) = 0,
\label{radialeqn}
\eeq
where $\mu$ is the ratio of the shell mass to the Planck mass, $\mu = m/m_p$. The general solution of the radial equation in 
\eqref{radialeqn} behaves as $e^{\pm (m-\cE)R}$ at large $R$, and can be expressed as a linear combination of Bessel functions 
of the first and second kind,
\beq
\psi(R) = \sqrt{R}\left[C_1 J_\sigma(-i\alpha R) + C_2 Y_\sigma(-i\alpha R)\right]
\eeq
where we let $\alpha = m-\cE > 0$ and $\sigma = \frac 12 \sqrt{1-\mu^4}$. Normalizability, according to \eqref{inprod}, 
requires $\Psi$ to fall off exponentially at infinity, with implies that $C_1 = i C_2$.  Thus $\phi(R)$ is Hankel's Bessel function 
of the third kind and the exact solution is
\beq
\Psi(\tau,R) = C e^{-i\cE\tau}\sqrt{R} H^{(2)}_\sigma(-i\alpha R)
\label{exactsol}
\eeq
where $C$ is an overall constant. As $R\rightarrow \infty$ the wave function falls off exponentially and, as $R\rightarrow 0$,
\beq
\Psi(\tau,R) \sim \left\{\begin{matrix}
C\sqrt{R}e^{-i\cE\tau}\left[\frac i\pi \Gamma(\sigma)\left(\frac{\alpha R}2\right)^{-\sigma} e^{\frac{i\pi\sigma}2}
+ \frac{(1-i\cot\pi\sigma)}{\Gamma(1+\sigma)} \left(\frac{\alpha R}2\right)^{\sigma}e^{-\frac{i\pi\sigma}2}\right],~~ \sigma\neq 0\cr\cr
\frac{2C}\pi \sqrt{R} \left[\gamma +\ln \left(\frac{\alpha R}2\right)\right],~~ \sigma = 0
\end{matrix}\right.
\label{nearzero}
\eeq
where $\gamma$ is Euler's constant. The behavior of these solutions near the center will depend on the mass ratio, $m/m_p=\mu$. If the 
shell mass is less than the Planck mass, $\mu < 1$, then $0\leq\sigma<1/2$ is real (we exclude the case $m=0$ because our construction 
is valid only for a timelike shell), but if the shell's rest mass is greater than the Planck mass, $\sigma$ is imaginary. 

Consider two stationary solutions, $\Phi_{\cE'}$ and $\Psi_\cE$, with energies $\cE'$ and $\cE$ respectively. 
The inner product \eqref{inprod} becomes 
\beq
\langle\Phi_{\cE'},\Psi_\cE\rangle = \frac 12\left[2m-(\cE+\cE')\right]e^{-i(\cE-\cE')\tau}\int_0^\infty dR~ \phi^*_{\cE'}\psi_\cE
\eeq
and by the equation of motion, we have 
\bea
\phi^*_{E'}\psi_\cE'' - \left((m-\cE)^2 - f^2\right)\phi^*_{\cE'}\psi_\cE &=& 0\cr\cr
\psi_{E}\phi^{*''}_{\cE'} - \left((m-\cE')^2 - f^2\right)\psi_\cE\phi^*_{\cE'} &=& 0
\eea
Subtracting the second from the first,
\beq
\phi^*_{\cE'}\psi_\cE'' - \psi_{\cE}\phi^{*''}_{\cE'} = (\phi^*_{\cE'}\overleftrightarrow{\partial_R}\psi_\cE)' = 
(\cE-\cE')(\cE+\cE'-2m)\phi^*_{\cE'}\psi_\cE 
\eeq
and it follows that the inner product is a boundary term,
\beq
\langle\Phi_{\cE'},\Psi_\cE\rangle =  \left.\frac{i J_R}{(\cE'-\cE)}\right|_0^\infty
\label{inprod2}
\eeq
where
\beq
J_R = -\frac i2 e^{-i(\cE-\cE')\tau}\phi^*_{\cE'}\overleftrightarrow{\partial_R}\psi_\cE
\eeq
is the radial component of the $U(1)$ current in \eqref{conscurtau}. The exponential fall off of our wave function at infinity 
ensures that $J_R$ vanishes there. The inner product therefore depends only on the value of the radial current at the origin. 

To guarantee orthonormality of the wave functions, we must require that the inner product of two wave functions of different 
energies vanishes. In particular, this means that $J_R$ should vanish at the origin when $\cE\neq \cE'$. Evaluating $J_R$, using 
the behavior of the solutions in \eqref{nearzero}, we find 
\beq
J_R \sim \left\{\begin{matrix}
\frac{|C|^2}{\sin\pi\sigma} \left[\left(\frac{m-\cE}{m-\cE'}\right)^\sigma-\left(\frac{m-\cE'}{m-\cE}\right)^\sigma\right],~~ 
\sigma\neq 0\cr\cr
\frac{2|C|^2}{\pi^2}\left[4in\pi + 2 \ln \left(\frac{m-\cE'}{m-\cE}\right)\right],~~ \sigma=0
\end{matrix}\right.
\eeq
If $\sigma$ is real ($m\leq m_p$) $J_R$ does not vanish, therefore there is no orthogonal set of solutions in this case. However, if 
the mass of the shell is greater than the Planck mass then $\sigma$ is imaginary and letting $\sigma = i\beta$,
\beq
J_R \sim \frac{|C|^2}{\sinh\pi\beta} \left[\left(\frac{m-\cE}{m-\cE'}\right)^{i\beta}-\left(\frac{m-\cE}{m-\cE'}\right)^{-i\beta}\right]
\eeq
vanishes if 
\beq
\frac{m-\cE}{m-\cE'} = e^{n\pi/\beta}
\eeq
for any integer $n$. Now the energy operator commutes with the superhamiltonian and there is proof of the positivity of energy
in General Relativity. It is reasonable, therefore, to exclude negative energy states and take the ground state to have zero energy.
Then this amounts to an energy spectrum,
\beq
\cE_n = m \left(1-e^{-n\pi/\beta}\right),
\label{spectrum}
\eeq
where $n$ is a positive integer.

Thus, from the comoving observer's point of view, there is a quantum theory of the shell but only for masses larger than the Planck mass.
Both the wave function and the $U(1)$ charge current density vanish at the center and are well behaved everywhere. 
The energy spectrum is discrete and, near the center, each energy eigenfunction is a combination of an infalling wave and an outgoing wave,
\beq
\Psi_n(\tau,R) \sim \frac{ie^{-\frac{\pi\beta}2}}\pi\Gamma(i\beta) C\sqrt{R}\left[\underbrace{e^{-i(\cE_n\tau + \beta \ln 
\frac{\alpha_n R}2)}}_{\text{infalling}} + \frac{\pi}{\beta\Gamma^2(i\beta)\sinh\pi\beta} \underbrace{e^{-i(\cE_n\tau-\beta \ln 
\frac{\alpha_n R}2)}}_{\text{outgoing}}\right]
\label{nearzerowf}
\eeq
with only a relative phase shift that depends on $m/m_p$.

Positivity of the energy and a well behaved probability current are sufficient to build a separable Hilbert space for the collapsing shell 
and, so far, no additional conditions at the origin have been explicitly imposed on the wave functions. The eigenfunctions in \cite{hajicek92a} 
were interpreted as solutions of a {\it classical} field by analogy with scalar electrodynamics and the energy-momentum of this
{\it classical} field was also required to strictly vanish at the center. We will now show that, ff $\Psi_n(\tau,R)$ in \eqref{exactsol} is 
treated as a classical field, the energy-momentum is ill defined at the center.

The wave equation is derivable from the action
\beq
S = \int d^2\xi \left[(i\partial_\tau + m)\Psi^*(-i\partial_\tau + m)\Psi + \partial_R\Psi^*\partial_R \Psi - f^2|\Psi|^2\right]
\eeq
and translation invariance leads to a conserved stress energy tensor (density)
\beq
{\Theta^\mu}_\nu = L \delta^\mu_\nu - \frac{\partial L}{\partial(\partial_\mu\Psi^*)}\partial_\nu\Psi -\frac{\partial L}
{\partial(\partial_\mu\Psi)}\partial_\nu\Psi^*.
\eeq
We focus our attention on the energy-momentum current, $P^\mu = {\Theta^\mu}_\tau$. Generalizing to bilinear currents, we find 
the following expressions for its components (taking into account that ours is an elliptic Klein-Gordon equation)
\bea
P^\tau(\Phi,\Psi) &=& -\dot\Phi^*\dot \Psi + \Phi^{*'}\Psi' + (m^2-f^2)\Phi^*\Psi\cr\cr
P^R(\Phi,\Psi) &=& -\Phi^{*'}\dot\Psi - \dot\Phi^*\Psi'.
\eea
or, for our stationary states,
\bea
P^\tau &=& e^{-i(\cE-\cE')\tau}\left[i(\cE+\cE') \phi^*_{\cE'}\psi_\cE + \left(\phi^*_{\cE'} \psi_\cE'+\phi^{*'}_{\cE'} 
\psi_\cE\right)'\right]\cr\cr
P^R &=& -ie^{-i(\cE-\cE')\tau}\left(\cE'\phi^*_{\cE'} \psi_\cE' - \cE\phi^{*'}_{\cE'} \psi_\cE\right)
\eea
Consider only the case $m>m_p$, for which a well behaved $U(1)$ current exists. For the states in \eqref{exactsol} and \eqref{spectrum}, 
we find that near $R=0$,
\beq
P^R \sim \frac{e^{-\pi\beta}(\cE-\cE')}{\pi\beta\sinh\pi\beta} \left[1+ \sqrt{1+4\beta^2}
\cos\left(2\beta\ln\left[\frac{mR}2\right] + \tan^{-1}2\beta - 2\varphi_\beta\right)\right]
\eeq
where $\varphi_\beta$ is the phase of $\Gamma(i\beta)$. The radial momentum density, $P^R$, oscillates with infinite frequency in 
the limit as $R\rightarrow 0$ but it is finite as the center is approached. On the other hand, the energy density
\beq
P^\tau \sim \frac{e^{-\pi\beta}}{\pi\beta\sinh\pi\beta} \frac{\sqrt{1+4\beta^2}}R \sin\left(2\beta\ln\left[\frac{mR}2\right] + 
\tan^{-1}2\beta - 2\varphi_\beta\right),
\eeq
diverges as $1/R$ in this limit.

What is the role of the energy-momentum current in this quantum mechanical model? Requiring it to vanish at the center serves not to 
completely define the Hilbert space but to select a subset of an otherwise completely well defined system. In general, doing so would 
raise the possibility that the selected subset of states is incomplete under the inner product. Moreover, changing the status of the 
wave function to that of a classical field was justified in \cite{hajicek92a} via a formal analogy with scalar electrodynamics. But 
our starting point is a conservation law that was obtained via the junction conditions, not a fundamental action principle, and
attempts at recovering the shell conservation law via an action principle from a fundamental theory have not succeeded in recovering 
\eqref{genconst} \cite{hajicek01,louko98}. The additional conditions at the origin may also exclude important states, such as those 
representing collapse to a black hole or naked singularity. For the proper time observer, there are no states that can satisfy this 
condition but for the reasons just stated, we do not consider this a problem.

\section{Comparison of the Quantum Descriptions}

The results of the previous section contrast with and complement the results of \cite{hajicek92a}, where the interior observer only
finds solutions for shells of mass less than the Planck mass. For the interior observer, the Wheeler-Dewitt equation is hyperbolic
and the inner product is positive semi-definite only on positive energy states. As mentioned in the previous section, in
\cite{hajicek92a} the wave functions were also interpreted as a classical field and the classical field was required to carry no
energy and momentum to the center. This was unnecessary for the construction of the states themselves but considered to be a
reasonable physical requirement based on the similarity of this system with scalar electrodynamics. To compare the quantum description
of an observer in the interior with that of a comoving observer we must ask what are the states for the observer in the interior {\it
had these additional conditions at the center not been imposed}. It turns out that the only difference is a doubling of the bound
eigenstates, when the shell rest mass is less than the Planck mass.

The radial equation for positive energy stationary states reads,
\beq
\psi'' + \left[(E^2-m^2) + \frac{\mu^2 E}R + \frac{\mu^4}{4R^2}\right]\psi = 0,
\eeq
and one can show, as in section III, that the charge form bears the same relationship
to the radial charge current as \eqref{inprod2} and that the radial charge current does not vanish at $R=0$ when $\mu>1$ for two states
with different energies. Therefore there are no solutions when $m>m_p$. When $\mu < 1$, the radial charge current can be made to vanish
and orthogonal states can be defined. Scattering states are given by the Kummer function as indicated in \cite{hajicek92a} and this
is unaffected by imposing the additional requirement that the energy-momentum vanishes at the center. Bound states can be given in terms
of the confluent hypergeometric function. However, without also requiring that the energy-momentum vanishes at $R=0$, we obtain
\beq
\Psi^\pm_n(\tau,R) = C R^{\frac 12\pm\sigma} e^{-\alpha^\pm_n R}U(-n, 1\pm 2\sigma,2\alpha^\pm_n~R)
\eeq
where $U(a,b,x)$ is the confluent hypergeometric function, $n$ is a whole number, $\alpha^\pm_n = \sqrt{m^2-E_n^{\pm 2}}$, $\sigma=\frac 12 
\sqrt{1-\mu^4}$ and $E^\pm_n$ is given by
\beq
E^\pm_n = \frac{2 m \left(\lambda_\pm + n\right)}{\sqrt{\mu^4 + 4(\lambda_\pm +n)^2}}
\eeq
where $\lambda_\pm =\frac 12\left(1\pm \sigma\right)$. The subset $\{\psi^-_n\}$ is eliminated if the classical field energy-momentum is also 
required to vanish at the center, but then completeness of the subset $\{\psi^+_n\}$ must be explicitly verified.

In the proper time description, the Wheeler-DeWitt equation is elliptic and the inner product is positive semi-definite for all energies
less than the shell's rest mass. The comoving observer finds no solutions when $\mu<1$ but, when $\mu>1$, the solutions are given by
Hankel's function
\beq
\Psi_n(\tau,R) = C e^{-i\cE_n\tau} \sqrt{R} H_{i\beta}^{(2)}(-i\alpha_n R)
\eeq
where now $\alpha_n = m-\cE_n >0$ and, assuming a ground state of zero energy,
\beq
\cE_n = m(1-e^{-n\pi/\beta}).
\eeq
where $\beta = \frac 12\sqrt{\mu^4-1}$. There are no states for which the classical field energy-momentum vanishes at the origin.

\section{Concluding Remarks}

The self-gravitating shell provides a remarkably simple example of a quantum gravitational system that can be solved exactly in two
different time coordinates and compared. In this paper we have quantized the shell in comoving time and compared the result with its 
quantization in interior, Minkowski time \cite{hajicek92a}.

At a deeper level, we would like to compare what each quantization says about the geometry of spacetime. To do so one should be able
to reconstruct the geometry of spacetime from the quantum states. For the proper time quantization, because only bound states exist,
we can at least think in terms of an ``asymptotic'' geometry. But the ADM mass is now a symmetrized version of the operator
\beq
\hM = \widehat{\cH} f(R)(m-\widehat{\cH})^{-1},
\eeq
which is not diagonalized in the Hilbert space and so the asymptotic energy will be smeared. Its average value may be given meaning 
via the Klein-Gordon product. This would also be true of the states in \cite{hajicek92a} (for shell masses smaller than the Planck mass) 
in approximately shell free regions created by constructing localized wave packets. In both quantizations, the other two time coordinates 
will be functions of the phase space variables as determined by \eqref{timeders}. On the quantum level they are operator valued and one 
can speak about time intervals in the other two regions only in terms of averages. The smeared ADM mass and time intervals imply that 
one must always deal with fuzzy local geometries in approximately shell-free regions in the interior and exterior. While this is not
surprising and these issues are present in any theory of quantum gravity, the positive semi-definite inner product available in the proper 
time formulation can be used to unambiguously evaluate the average values and quantify the fluctuations in the local geometry.  In this 
sense, the proper time quantization provides the simplest setting in which these questions can be meaningfully addressed.

We conclude by elaborating on the surprising structural similarity between the proper time Hamiltonian for the shell and its counterpart
for a dust ball. This is surprising because the quantum theory of the shell is derived from the junction conditions whereas the 
Hamiltonian in \cite{vaz01} was obtained from a canonical reduction of the full Einstein-Dust system. We can show that the structural 
similarity between the two runs deeper than \eqref{superh1}. Consider a shell that is collapsing onto some spherical object such as a 
pre-existing star or black hole. The setup of sections II and III can be used to good effect in this case: the constraint is given by 
\eqref{genconst}. In analogy with the shell collapsing in a vacuum, one can associate $\Delta M$ with the energy that is responsible for 
evolving the system in the internal time, $t_-$. Since neither the time nor the coefficients of the external Schwarzschild metric will 
play a role in the following, we let $t=t_-$ and $B^- = B$. Then using the relations \eqref{timeders} we have 
\beq
\Delta M = E = \frac{m B^{3/2}}{\sqrt{B^2-R_t^2}} - \frac{Gm^2}{2R}
\eeq
As before, one can determine a Hamiltonian for the evolution in $t$,
\beq
H = \sqrt{m^2 B + B^2p^2} -\frac{Gm^2}{2R},
\eeq
where
\beq
p = \frac{m R_t}{\sqrt{B^3-BR_t^2}},
\eeq
and an action
\beq
S = \int dt \left[-m\sqrt{B-\frac{R_t^2}B} + \frac{Gm^2}{2R}\right].
\eeq
The action in proper time is then recovered by using the relations \eqref{timeders}. We find
\beq
S= \int d\tau\left[-m + \frac{Gm^2}{2R}\sqrt{B+R_\tau^2}\right]
\eeq
and from from here derive the Hamiltonian for the evolution of the shell in proper time,
\beq
\cH = - P_\tau = m - \sqrt{\frac{f^2}B - B P^2},
\eeq
where 
\beq
P = \frac{mR_\tau}{B\sqrt{B+R_\tau^2}}.
\eeq
Thus we can base a quantum theory of the shell on the superhamiltonian
\beq
h = (P_\tau + m)^2 + B P^2 -\frac{f^2}B
\eeq
by requiring 
\beq
\widehat{h}\Psi(\tau,R) = \left[(\hP_\tau + m)^2 + B \hP^2 -\frac{f^2}B\right]\Psi(\tau,R) = 0
\eeq
Once again, this equation has exactly the same structure as the superhamiltonian for a dust ball, if the dust ball is thought of as
made up of a sequence of shells labeled by the LTB radial coordinate. The mass inside the shell, that appears in $B$, gets replaced
by the Misner-Sharpe mass function up to the radial coordinate of the shell in question. We will report on the analysis of this system
in a future publication.
\bigskip

\noindent In memory of my advisor and friend, Professor Gabor Domokos.

\end{document}